\documentclass[preprint,aps,pra,showpacs,floats,superscriptaddress]{revtex4}
\usepackage[dvips]{graphicx}
\usepackage{amsmath,bm}
\usepackage{psfrag}

\begin{document}

\title{Morphologies and kinetics of a dewetting ultrathin solid film}

\author{M.~Khenner}
\affiliation{Department of Mathematics, State University of New
York at Buffalo, Buffalo, NY 14260, USA}

\newcommand{\Section}[1]{\setcounter{equation}{0} \section{#1}}
\newcommand{\rf}[1]{(\ref{#1})}
\newcommand{\beq}[1]{ \begin{equation}\label{#1} }
\newcommand{\eeq}{\end{equation} }

\begin{abstract}

The surface evolution model based on geometric partial differential equation
is used to numerically study the kinetics of dewetting and dynamic morphologies for the
localized pinhole defect in the surface of the ultrathin solid film with the strongly anisotropic
surface energy. Depending on parameters such as the initial depth and width of the pinole,
the strength of the attractive substrate potential and the strength of the surface energy anisotropy,
the pinhole may either extend to the substrate and thus rupture the film, or evolve to the quasiequilibrium
shape while the rest of the film surface undergoes phase separation into a hill-and-valley structure
followed by coarsening. Overhanging (non-graph) morphologies are possible for deep, narrow (slit-like) pinholes.
\end{abstract}
\pacs{68.55.-a}

\date{\today}
\maketitle

\begin{center}
{\bf I. INTRODUCTION}
\end{center}

Recent experiments \cite{YZSLLHL,SECS} with the sub-10nm silicon-on-insulator films at 800-900 $^\circ$C 
demonstrate Si film dewetting caused by the long-range film-substrate interactions 
(which are also called wetting interactions). The mass transport in this system is by thermally-activated surface diffusion, and there is no stress 
at the film-substrate interface due to an absence of a lattice mismatch between the film and the SiO$_2$ substrate.
Dewetting starts at randomly distributed pinhole defects
in the Si planar surface. 
The pinholes may exist prior to the annealing, or they form shortly after the temperature is raised.
Conditions favoring pinhole deepening over contraction, where the latter is caused by minimization of the surface area due to 
the mean surface energy (tension), and 
the kinetics of the pinhole and its dynamic shape are not presently known. 
Besides, in contrast to liquids, the surface energy of solid surfaces is strongly anisotropic,
leading to missing orientations in the dynamic or equilibrium surface shape \cite{Wulff}-\cite{Bonzel_physrep} and faceting instability \cite{LiuMetiu} - \cite{WORD}.
These factors are certain to affect the dynamics of the pinhole and moreover, 
through the nonlinear competition with the attractive wetting potential
may lead to the emergence of an equilibria and thus to the suppression of the film dewetting and rupture.

In Ref. \cite{Dwt1}, following Ref. \cite{GLSD}, the partial differential equation (PDE)-based model is developed which allows to predict the 
wavelength of the fastest growing cosine-like perturbation of the film surface
(also called the normal perturbation), assuming
surface diffusion and the two-layer wetting potential \cite{ORS}-\cite{BrianWet}.
The model also enables computation of the dynamical (and probably faceted) morphologies.
Such computations are performed for the normal perturbation and they
demonstrate the stabilizing impact of the surface energy anisotropy on dewetting dynamics.
The model can in principle support any reasonable form of the wetting potential \cite{SZ,Chiu} 
but the corresponding contribution to the governing PDE for the film thickness must be rederived.

In this paper, using the model of Ref. \cite{Dwt1}, the kinetics and morphologies are computed systematically
for the \textit{localized} surface defect from the full nonlinear PDE. As has been made clear above, real surface 
defects are necessarily
localized. We compute for different widths and depths of the pinhole and, for all other model parameters 
fixed, observe very different dynamics and dewetting outcomes. 
We also relax the assumption made in Ref. \cite{Dwt1} that the surface height above the substrate 
is described by a function $h(x,t)$ (i.e., a one-dimensional (1D) surface is nonoverhanging) and reformulate the model in terms of two
parametric PDEs. This allows to compute, say, beyond the surface phase separation \cite{LiuMetiu,SG} into orientations $0^\circ$ and 
$90^\circ$ for some surface energy anisotropies. 
Asymmetric morphologies and different kinetics may arise when the direction of the maximum surface energy is not the reference direction
for the shape evolution (i.e., for instance, the $z$-axis perpendicular to the substrate), which is often the case.
%Overhanging morphologies may arise when the initial surface is 
%misoriented  with respect to [01] crystalline direction, which is often the case. 
Thus we incorporate such misorientation in the model. Note that parametric formulations of the 
geometric surface evolution laws are common, see for instance Refs. \cite{BKL,HLS,KBM}.

\begin{center}
{\bf II. PROBLEM STATEMENT}
\end{center}

%\underline {\emph{The Model.}}$\;$ 
A 2D film with the free 1D parametric surface $\Upsilon(x(u,t),z(u,t))$ is assumed, where
$x$ and $z$ are the Cartesian coordinates of a point on a surface, $t$ is time and $u$ is the parameter along the surface.
The origin of a Cartesian reference frame is on the substrate, and along the substrate ($x$-direction, or the
[10] crystalline direction) the film is 
assumed infinite. The $z$-axis is  along the [01] crystalline direction, which is normal
to the substrate.
Marker particles are used to track the surface evolution (see Ref. \cite{Trygg}, for instance) thus $x$ and $z$ 
in fact represent the coordinates of 
a marker particle, which are governed by two coupled parabolic PDEs \cite{Sethian1,Sethian2,BKL,HLS}:
\begin{subequations}
\label{base_eq}
\begin{eqnarray}
x_t &=& V \frac{1}{g}z_u,\\
z_t &=& - V \frac{1}{g}x_u.
\end{eqnarray}
\end{subequations}
Here the subscripts $t$ and $u$ denote differentiation,
$V$ is the normal velocity of the surface which incorporates the physics of the problem,
and $g=ds/du=\sqrt{x_u^2+z_u^2}$ is the metric function (where $s$ is the arclength). 
%The misorientation equals to zero means
%that the initial surface is prepared perpendicular to the $z$-axis.

The normal velocity of the surface is due to gradients of the surface chemical potential $\mu$, which drive the mass flux of
adatoms along the surface. In other words, redistribution of adatoms along the surface changes its shape, which is equivalent to the surface moving in the normal direction \cite{MULLINS5759,MULLINS_Metals,TCH94}. 
The chemical potential is the sum of two contributions,
$\mu = \mu^{(\kappa)}+\mu^{(w)}$, where $\mu^{(\kappa)}$ is the regular contribution due to the surface mean curvature 
$\kappa$, and $\mu^{(w)}$ is the wetting chemical potential \cite{GLSD,LGDV,Dwt1}:
\begin{equation}
\mu^{(w)} = \Omega \left(\gamma_p(\theta) - \gamma_S\right)\frac{\exp{\left(-z/\ell\right)}}{\ell} \cos{\theta},\quad z>0.
\label{1.4e}
\end{equation}
Here $\Omega$ is the atomic volume, $\theta$ is the angle that the unit surface normal makes with the [01] crystalline direction, $\gamma_S=const.$ is the surface energy of the substrate in the absence
of the film, $\ell$ is the characteristic wetting length, and $\gamma_p(\theta)$ is the primary
part of the anisotropic surface energy of the film, i.e. for typical four-fold anisotropy
\begin{equation}
\gamma(\theta) = \gamma_0 (1+\epsilon_\gamma \cos{4(\theta+\beta)}) + \frac{\delta}{2}\kappa^2 \equiv 
\gamma_p(\theta) + \frac{\delta}{2}\kappa^2.
\label{1.2}
\end{equation}
In Eq. \rf{1.2} $\gamma_0$ is the mean value of the surface energy, $\epsilon_\gamma$ determines the degree of anisotropy, $\beta$ is the misorientation angle 
%(where $\beta > 0$ signals that the initial surface has its
%normal slightly misoriented away from the [01] direction), 
and $\delta$ is the small non-negative regularization parameter having units of energy. 
The $\delta$-term in Eq. \rf{1.2} makes the evolution equations \rf{base_eq} mathematically well-posed for strong 
anisotropy \cite{AG} - \cite{Brian_regular}, \cite{LiuMetiu,SG,GoDaNe98}. 
(The anisotropy is weak when $0<\epsilon_\gamma<1/15$ and strong when $\epsilon_\gamma\ge 1/15$.
$\delta=0$ in the former case.) The surface energy has a maximum at $4(\theta+\beta)=0$, i.e. at $\theta=-\beta$.
For $\beta=0$, this direction is the $z$-axis.

The curvature contribution to $\mu$ is \cite{GLSD,LGDV,Dwt1}
\begin{equation}
\mu^{(\kappa)} = 
\Omega\left[\left(\gamma_p+\frac{\partial^2\gamma_p}{\partial \theta^2}\right)
\left(1-\exp{\left(-z/\ell\right)}\right)\kappa + \gamma_S\exp{\left(-z/\ell\right)}\kappa - \frac{\delta}{\gamma_0}\left(\frac{\kappa^3}{2}+\kappa_{ss}\right)\right], 
\label{1.4b}
\end{equation}
where the subscript $s$ denotes differentiation with respect to the arclength.
If the wetting potential is zero ($z/\ell \rightarrow \infty$), this reduces to familiar strongly anisotropic form \cite{LiuMetiu,SG,GoDaNe98}.

Finally, 
\begin{equation}
\kappa = g^{-3}\left(z_{uu}x_u-x_{uu}z_u\right)
\label{1.4a}
\end{equation}
and
\begin{equation}
V= \frac{D\nu}{kT}\left(\mu^{(\kappa)}_{ss} + \mu^{(w)}_{ss}\right),
\label{1.4c}
\end{equation}
where $D$ is the adatoms diffusivity, $\nu$ the adatoms surface density, $k$ the Boltzmann constant and
$T$ the absolute temperature.
The only differences of this formulation from the formulation in Ref. \cite{Dwt1}, except for accounting
for 
%the initial crystal surface 
the surface energy misorientation in Eq. \rf{1.2} and the parametric representation,
are that in $\mu^{(\kappa)}$, the wetting (exponential) contributions
are accounted for in full (that is, the approximation in the form of averaging across the film thickness is 
not employed) and 
the regularization term is not
included in $\mu^{(w)}$. ($\mu^{(w)}$ does not contain the 
surface stiffness $\gamma+\gamma_{\theta\theta}$ and thus it does not make the PDE ill-posed for strong anisotropy.
Besides, the regularization contribution to $\mu^{(w)}$ 
is vanishingly small for large surface slopes due to its proportionality to $h_x^{-7}$ \cite{Dwt1}).

To nondimensionalize the problem,  the thickness of the planar undisturbed film, $h_0$, is chosen as the length scale,
and $h_0^2/D$ as the time scale. Also, let $r = \ell/h_0$. 
The dimensionless problem is comprised of Eq. \rf{1.4a} and Eqs. \rf{base_eq} (where differentiations
are with respect to the dimensionless variables), and where 
\begin{subequations}
\label{ndim_eq}
\begin{eqnarray}
V &=& B\left(\mu^{(\kappa)}_{ss} + \mu^{(w)}_{ss}\right),\\
\mu^{(\kappa)} &=& 
\left(\gamma_p+\frac{\partial^2\gamma_p}{\partial \theta^2}\right)
\left(1-\exp{\left(-z/r\right)}\right)\kappa + \Gamma\exp{\left(-z/r\right)}\kappa - \Delta\left(\frac{\kappa^3}{2}+\kappa_{ss}\right),\\
\mu^{(w)} &=& \left(\gamma_p(\theta) - \Gamma\right)\frac{\exp{\left(-z/r\right)}}{r} \cos{\theta},\\
\gamma_p(\theta) &=& 1+\epsilon_\gamma \cos{4(\theta+\beta)}.
\end{eqnarray}
\end{subequations}
In Eqs. \rf{ndim_eq} $B = \Omega^2\nu \gamma_0/(kTh_0^2),\ \Gamma = \gamma_S/\gamma_0$ and 
$\Delta = \delta/(\gamma_0h_0^2)$.
For the computational method, using the relation between $s$ and $u$, 
\begin{equation}
\frac{\partial}{\partial s} = \frac{1}{g}\frac{\partial}{\partial u}\ ,
\label{partials}
\end{equation}
the problem is written entirely in terms of the independent variables $u$ and $t$ (not shown).

In the simulations reported below, the following values of the physical parameters are used:
$D=1.5\times 10^{-6}$ cm$^2$/s, $\Omega = 2\times10^{-23}$ cm$^3$, $\gamma_0 = 10^3$ erg/cm$^2$, 
$\gamma_S = 5\times10^2$ erg/cm$^2$, $\nu = 10^{15}$ cm$^{-2}$, $kT = 1.12\times10^{-13}$ erg, $h_0= 10^{-6}$ cm,
and $\delta = 5\times 10^{-12}$ erg. These values translate into $B=3.57\times10^{-3}$, $\Gamma = 0.5$ and 
$\Delta = 5\times 10^{-3}$ (or zero). In this paper we consider strong anisotropy, $\epsilon_\gamma > 1/15$ and 
$r=0.02,\ 0.1$. 

The initial condition in all runs is the Gaussian surface 
\begin{equation}
z(x,0) = 1-d\exp{\left[-\left(\frac{x-5}{w}\right)^2\right]},\quad 0\le x\le 10
\label{ini_cond}
\end{equation}
where $0<d<1$ and $w$ are
the depth and the ``width" of the pinhole at $t=0$, respectively (see Figure 4). The length of the computational domain 
equals to ten times the unperturbed film thickness, and the defect is positioned at the center of the domain.
We use values $d=0.5,\ 0.9$, which correspond to shallow and deep pinhole at $t=0$, respectively, 
and $w=0.15$ (narrow pinhole), $w=1$ (intermediate pinhole) and $w=2$ (wide pinhole).

The method of lines is used for the computation with the periodic boundary conditions at $x=0,\ 10$.
Eqs. \rf{base_eq} are discretized
by second-order finite differences on a spatially-uniform grid in $u$. 
The integration in time of the resulting coupled system of the 
ordinary differential equations
is done using the implicit Runge-Kutta method.
Initially $u\equiv x$, but periodically (usually after every few tens of the time steps) the surface is reparametrized
so that $u$ becomes the arclength, and the positions of the marker particles are recomputed accordingly.
This prevents marker particles from coming too close or too far apart in the course of surface evolution.

\begin{center}
{\bf III. RESULTS}
\end{center}

\begin{center}
{\bf A. Kinetics}
\end{center}

Figures 1 and 2 show the log-normal plots of the pinhole depth vs time, for $d=0.9$ and $d=0.5$, respectively.
%$\epsilon_\gamma = 1/12$.
$z_m$ is the height of the surface at the tip of the pinhole.
%(Note that at $t=0$, $d+z_m=1$.) 

Wide and intermediate deep pinholes dewet
but the depth of the narrow deep pinhole decreases until it reaches quasiequilibrium at $z=0.75$ (Figure 1). 
Quasiequilibrium means that $z_m$ (or, equivalently, the depth) changes very slow or not at all, while the rest of the shape 
may change relatively fast. 
Correspondingly, we will call the surface shape at the time when the quasiequilibrium depth is attained, the 
quasiequilibrium shape. (Again, this shape \textit{is} changing, but the height of its minimum point (the tip) is not
changing, or is changing very slow).
Also notice in Figures 1(a) and 1(b) that the growth rate at the rupture time
is finite for $r=0.1$ but infinite (or extremely large) for $r=0.02$, and the time to rupture is somewhat less
for $r=0.1$. However, from Figure 1(c), the time to reach quasiequilibrium is about ten times larger for $r=0.1$ 
than for $r=0.02$. 

As can be seen in Figure 2, only the wide shallow pinhole dewets, and only when
$r=0.1$. In all other cases of $w$ and $r$ (except $w=1,\ r=0.1$, shown by the solid
line in Figure 2(b)) the quasiequilibrium is achieved. 
For $w=1,\ r=0.1$, the depth is initially a non-monotonic function of time,
but after the transient phase it monotonically and slowly decreases without reaching the quasiequilibrium 
(we computed for $t\le 2\times 10^4$).
Also, one can see that for $r=0.02$ the depth at quasiequilibrium 
decreases as $w$ decreases, and it takes less time to reach quasiequilibrium as $w$ decreases. 

In Figure 3 the dewetting kinetics is compared for several misorientations and strengths of the anisotropy.
Shown is the case of the deep, wide pinhole and $r=0.1$. The time to rupture increases insignificantly 
with the decrease of 
anisotropy or with the increase of the misorientation angle. 
The faster dewetting for stronger anisotropy
here can be attributed to the initial faster shape changes due to larger 
gradients of $\mu^{(\kappa)}$,
i.e. before the surface orientation falls into an unstable (spinodal) range and faceting steps in, 
and to the proximity of the pinhole tip to the substrate. 
That the misorientation slows the kinetics 
is well-known, for instance Liu \& Metiu \cite{LiuMetiu} in their important study of faceting 
call the similar situation an ``off-critical quench", and find that at sufficiently large misorientations
the ``crystal surface will not phase-separate \textit{spontaneously}, but will have to overcome a finite free-energy
barrier".

\begin{center}
{\bf B. Morphologies}
\end{center}

Figure 4 shows the initial and the final surface shapes of the initially deep pinhole for the three values of $w$.
The kinetics of the corresponding \textit{dynamical} shapes is shown in Figures 1(a)-(c) by solid lines
and has been discussed above. 
Dewetting of the wide and intermediate pinholes proceeds through the extension of the tip of the pinhole until it reaches the
substrate at $57^\circ$.
The  quasiequilibrium shape for the narrow pinhole is very similar to the one 
shown in Figure 5(c). The latter shape is discussed in more details below.
While evolving from the initial slit-like shape to the quasiequilibrium shape, the surface of the narrow pinhole overhangs,  until it
slowly returns to the non-overhanging shape later (Figure 5).
In the time interval where the overhanging takes place, the surface slope is large and non-analytic.

Surface shapes for the initially shallow pinhole are shown in Figure 6. Characteristic of these shapes 
is emergence of the hill-and-valley structure \cite{HERRING1}, which becomes possible even when the film
dewets (case of the wide pinhole in Figure 6(a)) due to larger time required to reach the substrate in this case.
The angle at rupture is $90^\circ$. 
Figure 6(b) shows coarsening of the structure for the intermediate width case. As has been 
pointed out above at the discussion of Figure 2(b), in this case there is no quasiequilibrium (at least until
$t>2\times 10^4$), and the tip recedes 
towards the unperturbed height $h=1$. Note that the apparent recession rate is 
slower than the overall coarsening rate. $t=7240$
is the time when the two pyramidal structures
appear on the film surface. The slope of their 
walls is shown in the inset and it is almost constant in each of the four characteristic intervals of $x$, with each
interval corresponding to a facet. For $7240 < t \le 2\times 10^4$ the 
walls (facets) become more straight and the graph of $z_x(x)$ becomes nearly constant in each of the four 
characteristic intervals.   
Figure 6(c) shows the quasiequilibrium shape for the narrow pinhole.
Formation of the hill-and-valley structure followed by coarsening continues for $t>4$, as is evidenced
by the surface slope shown in the inset at the left, but the pinhole depth at $x=5$ is constant.
(In fact, the difference of depths at $t=197$ and at $t=4$ is 0.007.) Note that formation of the hill-and-valley 
structure, its coarsening
and slope selection have been the subject of many papers, see for instance Refs. \cite{LiuMetiu}-\cite{WORD},
\cite{Chiu}, \cite{GLSD,LGDV}, \cite{ORS}, \cite{SP}-\cite{Kohn}. (Refs. \cite{GLSD,LGDV,ORS} discuss impacts of wetting interactions). Since in this paper we are 
interested in characterizing dewetting and rupture, we do not further pursue that direction.

Finally, Figure 7 demonstrates impacts of misorientation ($\beta = 10^{\circ}$) on morphology.
As expected, asymmetrical shapes emerge for $\beta \neq 0^\circ$.
While the wide, shallow pinhole dewets, only its right sidewall undergoes the phase separation into a 
hill-and-valley structure.
Kinetics is very similar to the case $\beta=0$, see Figure 2(a).
%, being somewhat slower. 

\begin{center}
{\bf IV. DISCUSSION}
\end{center}

In this paper a fully nonlinear model is used to compute 
the complex scenarios of dewetting/equilibration 
for a localized pinhole defect in the surface of strongly anisotropic thin solid film, assuming the two-layer 
wetting potential and zero lattice mismatch with the substrate.

The computed dewetting kinetics can be explained at large using the magnitude  of the dewetting
factor. 
In Ref. \cite{Dwt1} it was shown that the dominant dewetting terms in the mass-conservation evolution PDE
are proportional to $\exp{\left(-z/r\right)}/r^2$ (for $r<1$). This is plotted in Figure 8.
One can see that for $r=0.1,\ 0.02$ the maximum of this factor occurs for small values of the film thickness.
This explains why all but one initial conditions in Figure 1 (deep pinhole) lead to pinhole depth increase until
the film dewets, 
while all but one initial conditions in Figure 2 lead to decreasing pinhole depth.
The dewetting factor can also explain the small (large) difference in characteristic time scales for 
$r=0.1$ and $r=0.02$ in Figure 1(a) (2(a)). Indeed, for the case of Figure 1(a) the factor is 36.8 ($r=0.1$) and 16.8
($r=0.02$) vs, respectively, 0.7 and $3\times 10^{-8}$ for the case of Figure 2(a).
Evolution of the film also depends strongly on the film shape and whether $\theta$ is in the 
unstable range. When it is, as in the cases shown in Figures 4-7 then, the faceting instability
is energetically more favorable than dewetting and most often the competition of the two processes causes the unusual
hill-and-valley structure where shallow pinhole remains despite the structure coarsening 
(which takes place separately at both shoulders of the pinhole). Note that the faceting instability 
seems to be always initiated where the surface changes from horizontal to a sidewall (regions A and B in
Figure 4).

We also point out that in our computations the pinhole tip is always nonfaceted at rupture (see Figures 6(a) and 7(a)). 
In fact, dewetting ceases if the facet spreads to the tip. 
Also, if the pinhole is not too deep (Figure 4), then the contact angle with the substrate at rupture is $90^\circ$.
In reality, after the contact this value decreases in order to minimize the total energy of the surface-substrate
system \cite{YZSLLHL}.

\begin{center}
{\bf ACKNOWLEDGMENT}
\end{center}

I thank Brian J. Spencer for the discussion and a few important ideas.

\newpage

\begin{center}
{\bf FIGURES CAPTIONS}
\end{center}
%\vspace*{1.0cm}\\

\noindent
Figure 1. Kinetics (rate) data for deep pinhole ($d=0.9$). $\epsilon_\gamma=1/12,\ \beta=0$.
Line slope equals the rate of the tip evolution. Solid line: $r=0.1$. Dash line: $r=0.02$.
(a) Wide pinhole ($w=2$). (b) Intermediate pinhole ($w=1$). (c) Narrow pinhole ($w=0.15$).
\vspace*{0.5cm}\\
\noindent
Figure 2. Same as Figure 1, but for shallow pinhole ($d=0.5$).
\vspace*{0.5cm}\\
\noindent
Figure 3. Kinetics data for deep, wide pinhole. $r=0.1$. Solid line: $\epsilon_\gamma=1/12,\ \beta=0$.
Dash line: $\epsilon_\gamma=1/12,\ \beta=10^\circ$. Dash-dot line: $\epsilon_\gamma=1/8,\ \beta=0$.
Dash-dash-dot-dot line: $\epsilon_\gamma=1/14,\ \beta=0$.
\vspace*{0.5cm}\\
\noindent
Figure 4. Surface morphologies for the deep pinhole. 
$r=0.1,\ \epsilon_\gamma=1/12,\ \beta=0$. The depth ($d$) and the width ($w$) of the
pinhole at $t=0$ are defined.
Solid, dash, dash-dot line: $w=2,\ 1,\ 0.15$, respectively. 
Inset: magnified view of dewetting region for $w=2,\ 1$.
\vspace*{0.5cm}\\
\noindent
Figure 5. Magnified view of the transient surface morphology and surface slope (inset) for the deep, narrow pinhole from Figure 4. Dash line: surface at $t=0$.
\vspace*{0.5cm}\\
\noindent
Figure 6.  Surface morphologies for the shallow pinhole. 
$r=0.1,\ \epsilon_\gamma=1/12,\ \beta=0$.
(a): $w=2$, (b): $w=1$, (c): $w=0.15$.
Dash line: surface at $t=0$. 
%\vspace*{0.5cm}\\
%\noindent
%Figure 7.  Surface morphologies. 
%$r=0.1,\ \epsilon_\gamma=1/12,\ \beta=10^\circ$.
%Dash line: surface at $t=0$. (a): $d=0.5,\ w=2$. (b): $d=0.9,\ w=0.15$.
\vspace*{0.5cm}\\
\noindent
Figure 7.  Surface morphology for the shallow, wide pinhole.
$r=0.1,\ \epsilon_\gamma=1/12,\ \beta=10^\circ$.
Dash line: surface at $t=0$. 
\vspace*{0.5cm}\\
\noindent
Figure 8. Contour plot of the dewetting factor $\exp{\left(-z/r\right)}/r^2$.

\newpage

\begin{figure}[!t]
\includegraphics[width=6.5in]{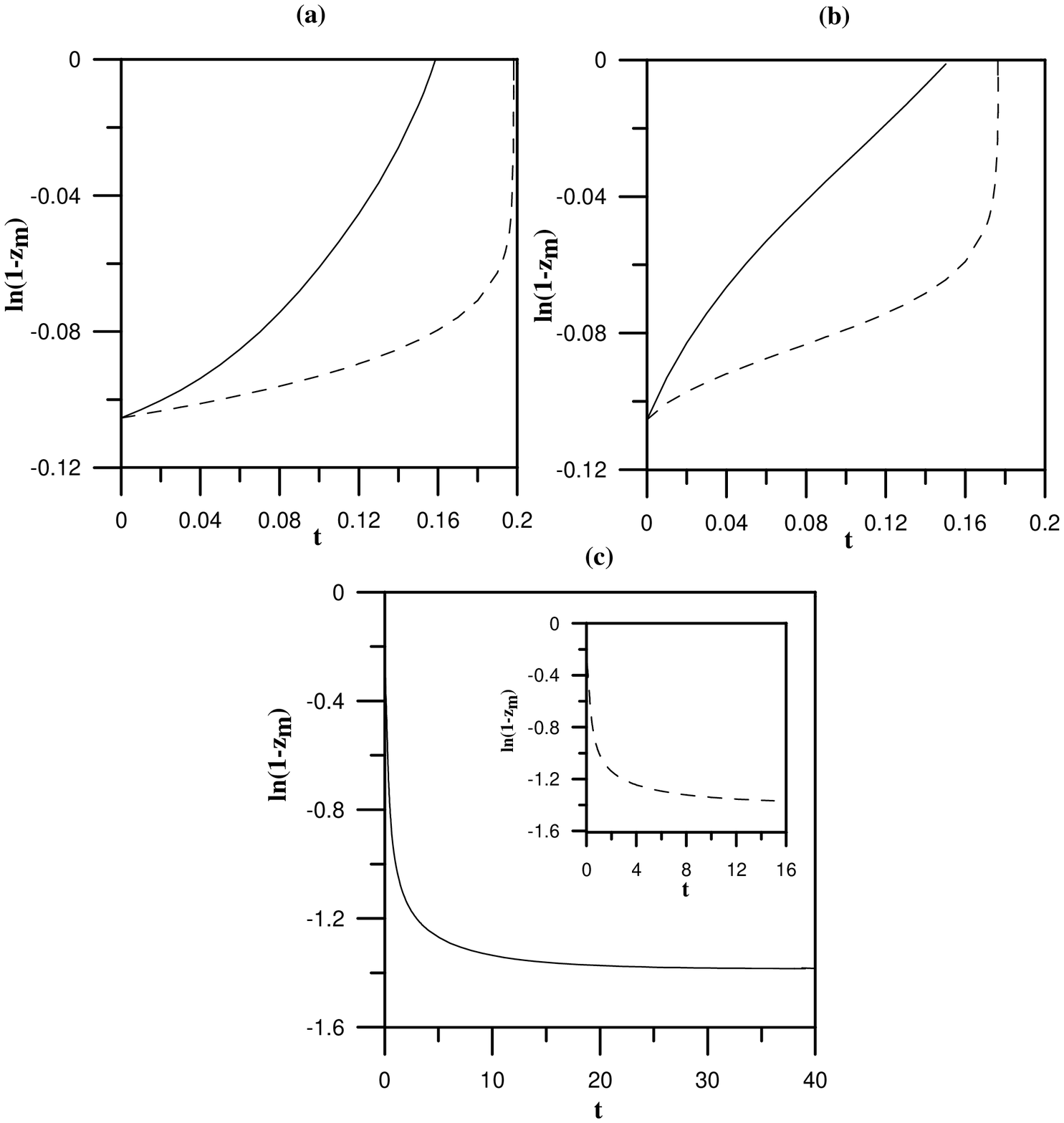}
\caption{} \label{fig:1}
\end{figure}

\newpage

\begin{figure}[!t]
\includegraphics[width=6.5in]{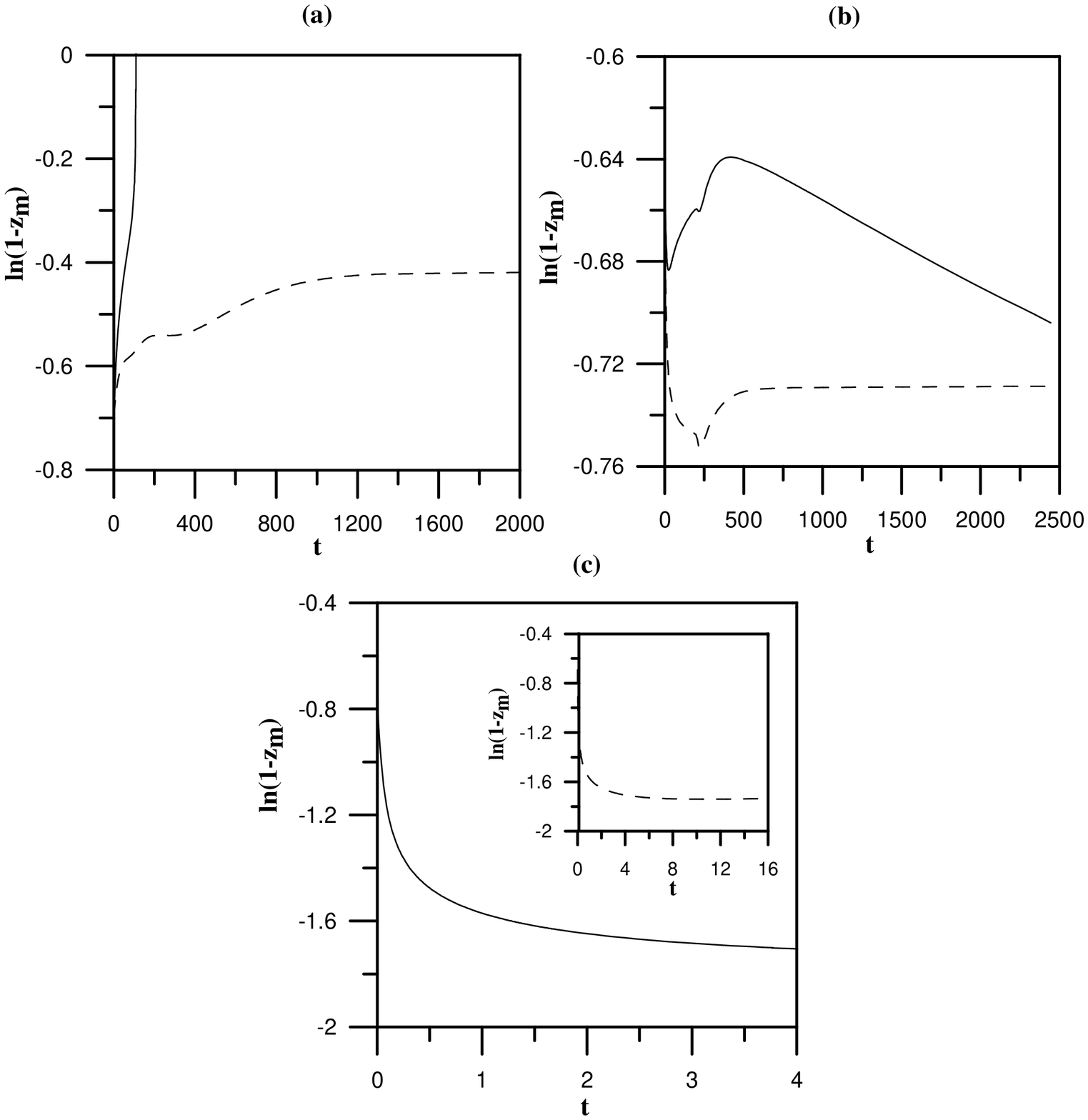}
\caption{} \label{fig:2}
\end{figure}

\newpage

\begin{figure}[!t]
\includegraphics[width=6.5in]{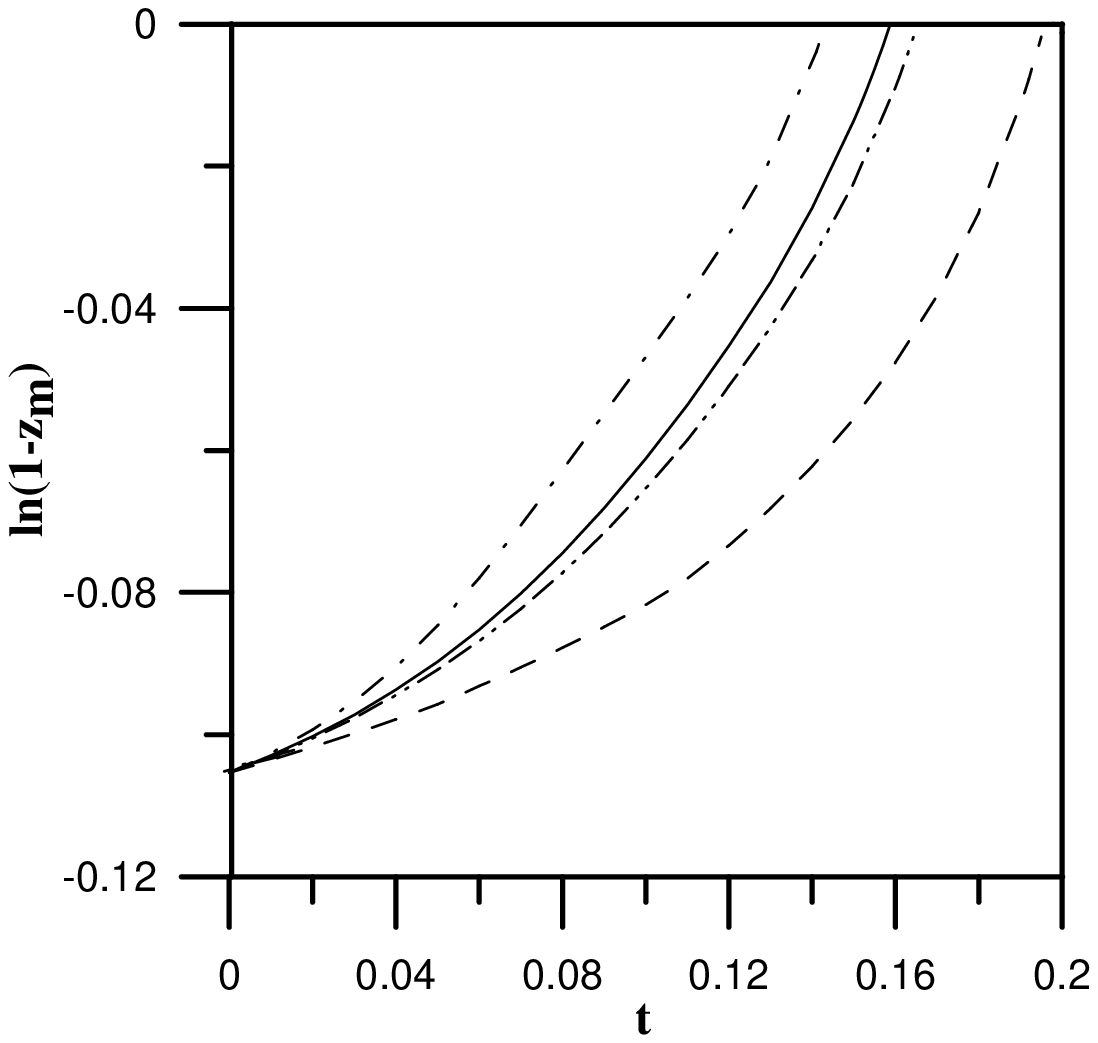}
\caption{} \label{fig:3}
\end{figure}

\newpage

\begin{figure}[!t]
\includegraphics[width=6.5in]{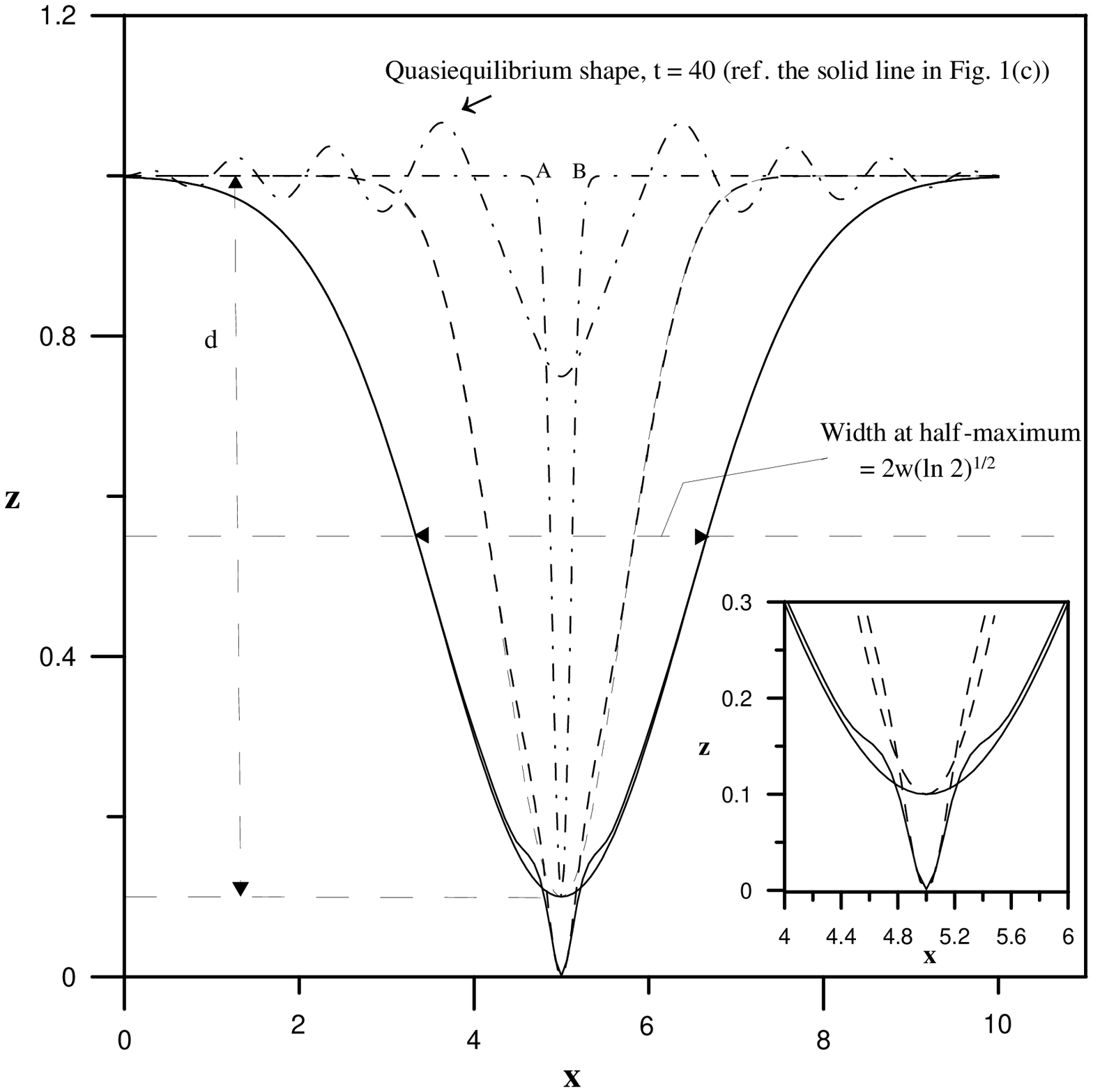}
\caption{} \label{fig:4}
\end{figure}

\newpage

\begin{figure}[!t]
\includegraphics[width=6.5in]{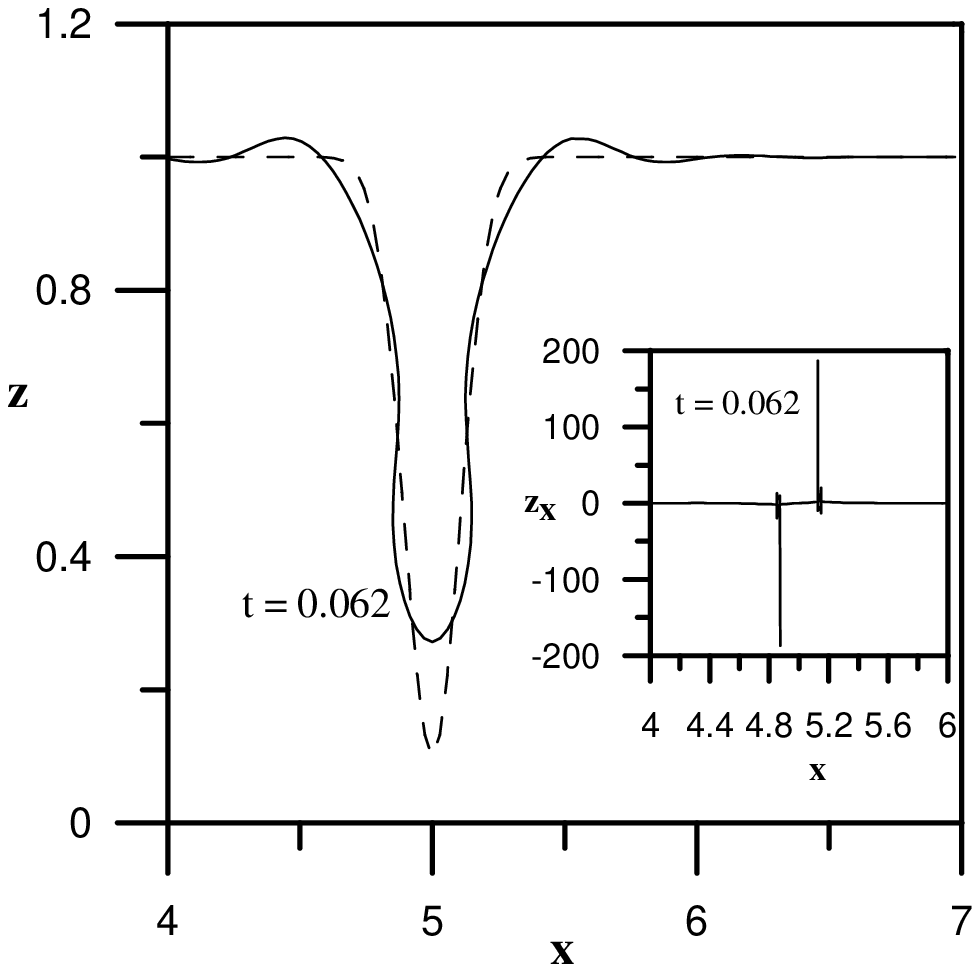}
\caption{} \label{fig:5}
\end{figure}

\newpage

\begin{figure}[!t]
\includegraphics[width=6.5in]{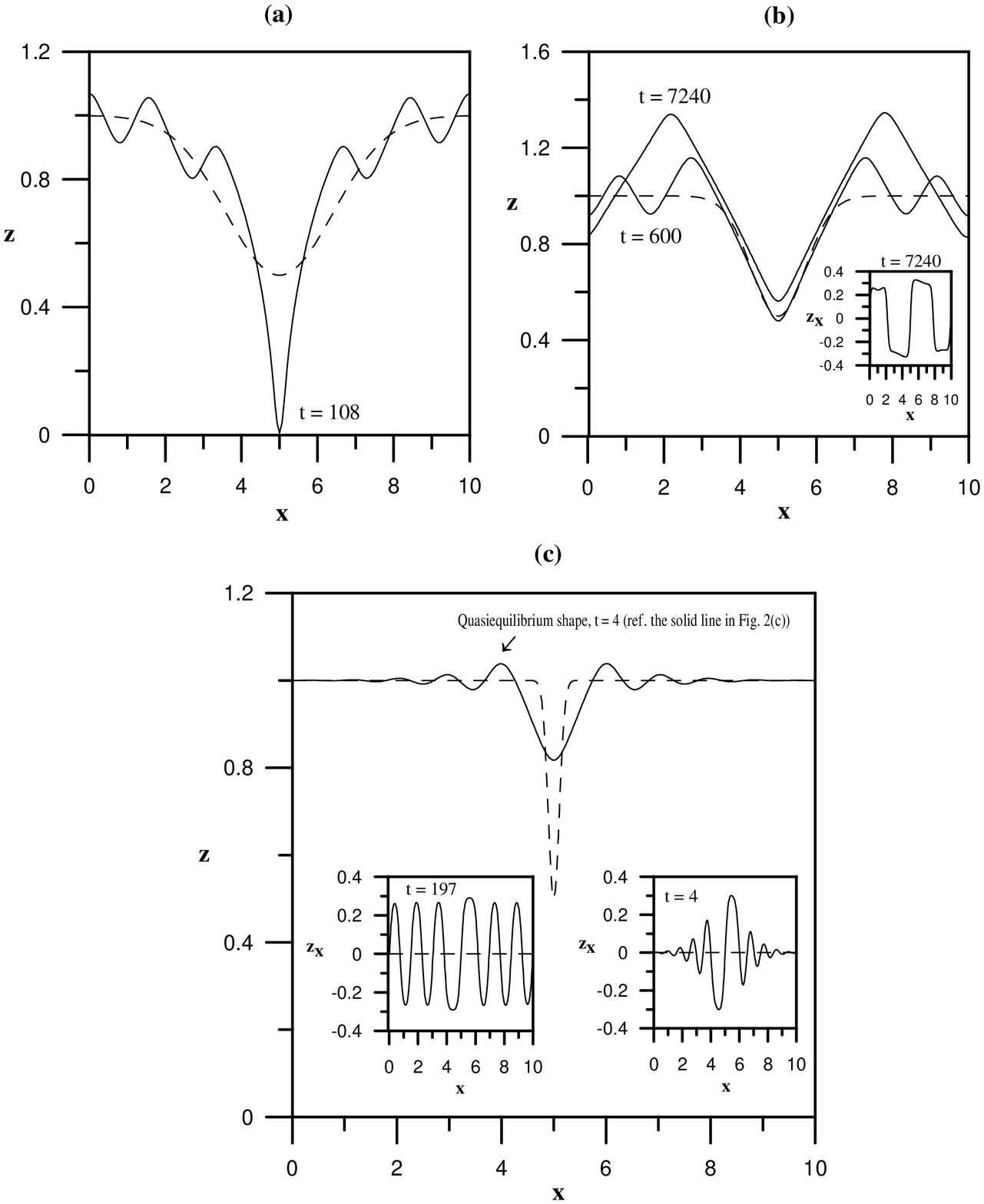}
\caption{} \label{fig:6}
\end{figure}

\newpage

\begin{figure}[!t]
\includegraphics[width=6.5in]{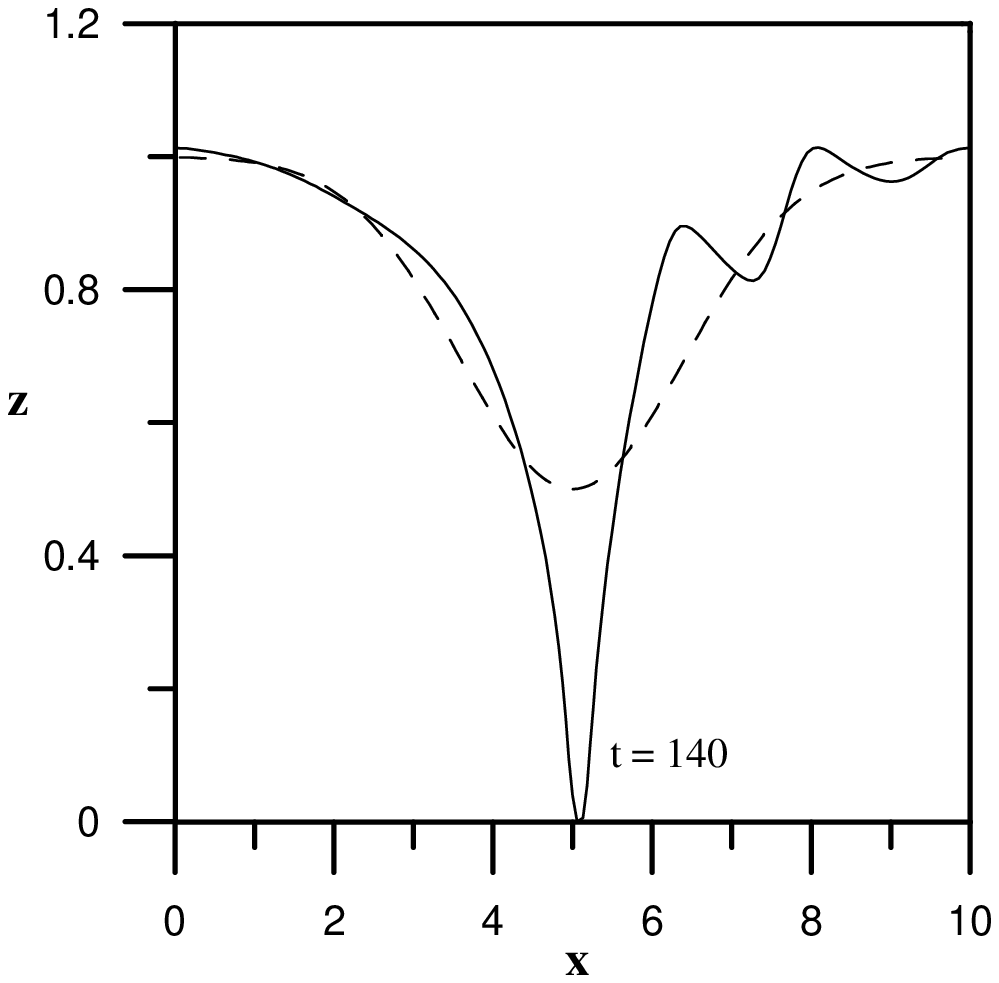}
\caption{} \label{fig:7}
\end{figure}

\newpage

\begin{figure}[!t]
\includegraphics[width=6.5in]{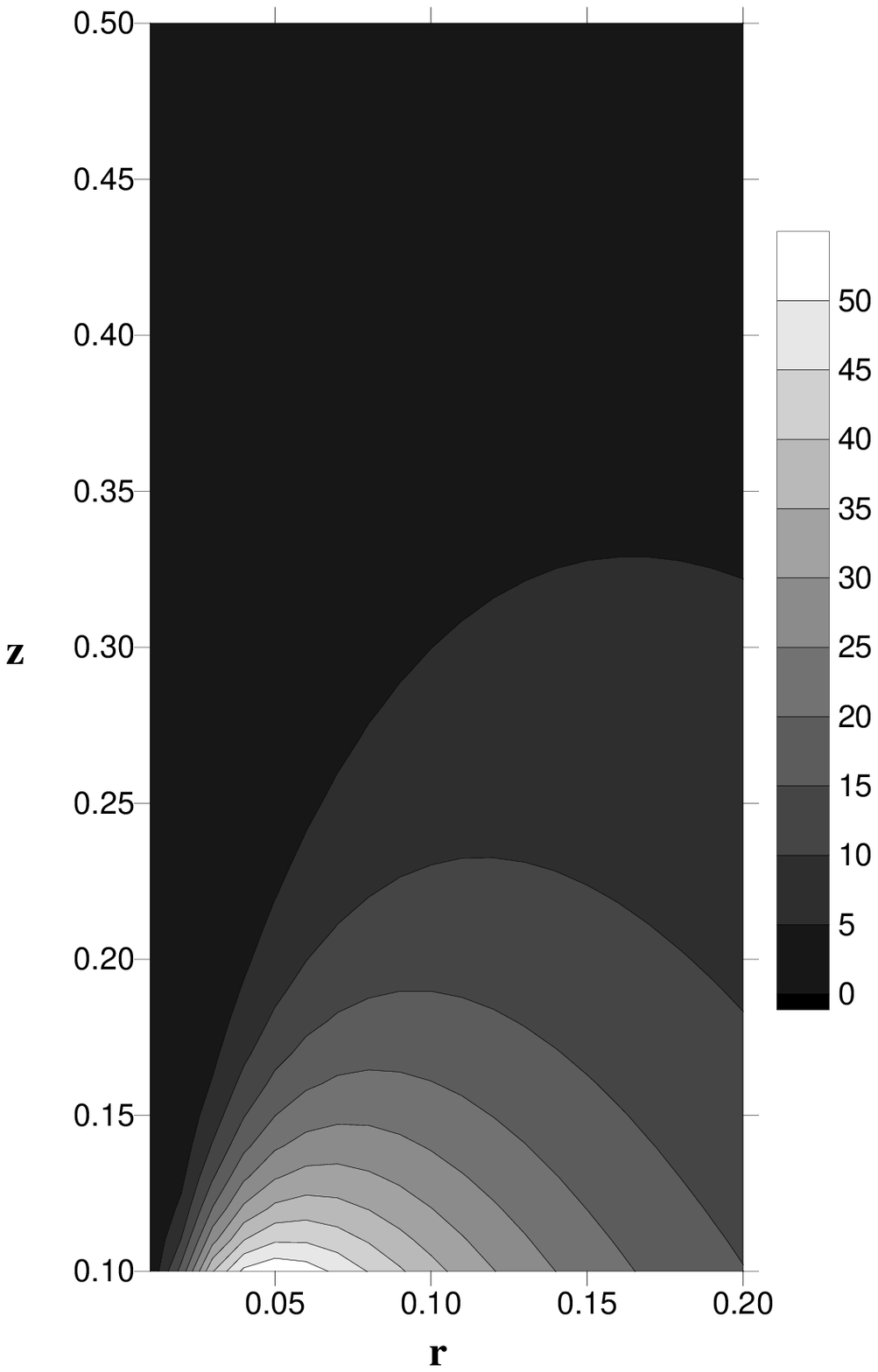}
\caption{} \label{fig:8}
\end{figure}

\end{document}